# The Stability of the Quantum Hall State at *v* = 1 in Bilayer Electron Systems

A.A. Vasilchenko

*National Research Tomsk State University, 634050 Tomsk, Russia*

Abstract

The ground states of electrons in two vertically coupled quantum dots in the presence of an external magnetic field have been studied within the density functional theory. A phase diagram of the transition to the quantum Hall state in double quantum dots has been plotted for the Landau level filling factor *v* = 1. It has been shown that the quantum Hall state can be stable for a zero tunnel splitting. The influence of the impurity potential on the stability condition for the quantum Hall state has been analyzed.

## 1. Introduction

Experimental studies [1 – 4] on the quantum Hall state (QHS) at a filling factor *v* = 1 in two-layer electron systems in GaAs showed a number of striking phenomena. If the distance *d* between the layers is large, then the layers behave independently at a filling factor *v* = 1/2 for each layer, and the quantum Hall effect is not observed. When *d* and the electron density decrease, a phase transition occurs and a Hall plateau appears at a filling factor *v* = 1. The peculiarity of this phenomenon is that the Hall plateau at *v* = 1 exists at a zero tunnel splitting. Provided that *d*/*L* < 2 (*L* is the magnetic length), the quantum Hall effect at *v* = 1 is observed at any values of the tunnel splitting [3]. Further experimental studies [4] showed that the critical ratio *d*/*L* is slightly less and a plateau is formed at *d*/*L* < 1.7. This phenomenon is often explained by a Bose condensate of interlayer excitons [5, 6]. However, the reason for this phenomenon has not been fully clarified.

A quantitative theory of the stability of QHS at *v* = 1 in a system of two vertically coupled quantum dots with a large number of *N*-electrons is developed in this work. The density functional theory is used to study the electronic properties of two vertically coupled two-dimensional quantum dots in the perpendicular magnetic field.

## 2. Theoretical model

Two identical quantum dots are separated by a barrier and are located at a distance $d$ from each other. The energy levels of symmetric and antisymmetric states are separated by a tunnel splitting $\Delta_{sas}$, isospin is defined as $I = (N_s - N_a)/2$ ($N_s$ ($N_a$) is the number of electrons in the subband of symmetric (antisymmetric) states). Electrons are confined inside quantum dots by parabolic potentials.

The effective system of atomic units is used, in which the energy is expressed in units of $Ry = e^2/(2\varepsilon a_B)$, and the length is expressed in units of $a_B = \varepsilon \hbar^2/(m_e e^2)$, where $m_e$ is the effective electron mass, $\varepsilon$ is the dielectric constant. The calculations are carried out for a GaAs quantum dot, for which $a_B = 9.8$ nm, $Ry = 5.9$ meV.

The total energy of two-dimensional electrons for two identical quantum dots can be written in the following form

$$E = T + \int (V_c(r,0) + V_c(r,d))n(r)dr + 2\int \varepsilon_x(n)n(r)dr - \sum_m \int (2\varepsilon_x(n_m) + V_{c,m}(r,0) + V_{c,m}(r,d))n_m(r)dr + 2\int V(r)n(r)dr - \Delta_{sas}(N_s - N_a)/2, \quad (1)$$

where $T$ is the kinetic energy of noninteracting electrons in a magnetic field $B$, which is given by the vector potential $A = B(-y/2, x/2, 0)$, $\Delta_{sas}$ is the tunnel splitting,

$$V_c(r,d) = 2\int \frac{n(r')dr'}{\sqrt{|r-r'|^2 + d^2}}, \quad V_{c,m}(r,d) = 2\int \frac{n_m(r')dr'}{\sqrt{|r-r'|^2 + d^2}}. \quad (2)$$

Expression (1) excludes the self-action of electrons, and the exchange energy in the local density approximation is used. For spin-polarized electrons, the exchange energy per electron has the form

$$\varepsilon_x(n) = \alpha n(r), \quad (3)$$

where $\alpha = -\sqrt{2\pi}\pi L$.

For spin-polarized electrons, the Kohn-Sham equations read

$$\{-\frac{\partial^2}{\partial r^2} - \frac{1}{r}\frac{\partial}{\partial r} + \frac{r^2}{4L^4} + \frac{m^2}{r^2} - \frac{m}{L^2} + V_{eff}(r)\}\psi_m(r) = E_m\psi_m(r), \quad (4)$$

with an effective single-particle potential:

$$V_{eff}(r) = V_c(r,0) + V_c(r,d) - V_{c,m}(r,0) - V_{c,m}(r,d) + 2\alpha(n(r) - n_m(r)) + V(r), \qquad (5)$$

$$V(r) = \frac{\omega_0^2}{4} r^2, \qquad (6)$$

where $m$ is angular momentum of an electron, $n_m(r) = |\psi_m(r)|^2 / 2$, $n(r) = \sum_{occ\ m} n_m(r)$.

## 3. Numerical results and discussion

In the case of identical quantum dots, the Schrödinger equation from the system of Kohn-Sham equations (4) – (6) is solved for electrons located in the same quantum dot. The total energy of electrons is calculated by formula (1). All calculations are performed for GaAs quantum dots. The comparison with the results of exact diagonalization [7] for quantum dots with $N = 3$ and 4 electrons showed that the energy difference is about 5%.

In QHS at $v = 1$ all electrons occupy only the energy level of the symmetric state and have the maximum isospin $I = N/2$ and the minimum total angular momentum $M = N(N-1)/2$, while all electrons in the angular momentum space have the configuration (0, 1, …, N-1). The transition from this state to a state with a higher angular momentum at high values of $\Delta_{sas}$ is studied The curves for $B_1(\omega_0)$ on Figure 1 show a transition $(0,1,...N-1) \to (1,2,...N)$, which is cloe to the transition from the state $v = 1$ to the state $v < 1$. Then, near the curve $B_1(\omega_0)$ (when $B < B_1$) the transition from QHS to the state with $N_a = 1$ is studied and the minimum tunnel splitting is found. Moreover, it is assumed that all electrons in the subband of symmetric state have a compact configuration in the angular momentum space. Phase diagram of such a transition is shown in Figure 2. It can be seen that at low magnetic fields (low electron densities) QHS is energetically favorable at any $\Delta_{sas}$ values. Note that the transition from QHS always occurs to the state with $N_a = 1$, and with a further increase of $B$ the $N_a$ value of increases. Such system behavior in the macroscopic limit will lead to the appearance of a Hall plateau at low electron densities in a structure with zero tunnel splitting.

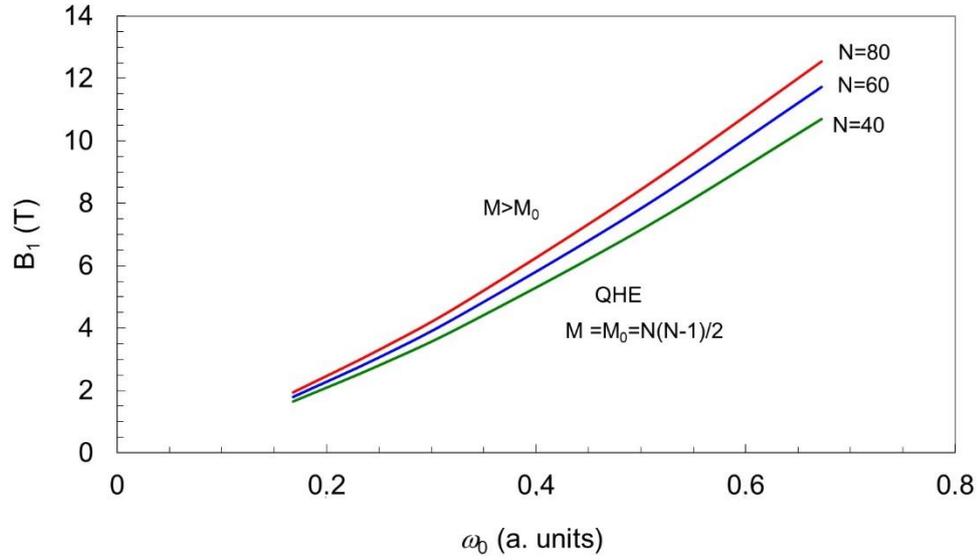

Figure 1. Transition from the state with the configuration of electrons (0, 1,…, $N$-1) to the state with configuration (1, 2,…, $N$); $d$ =1.

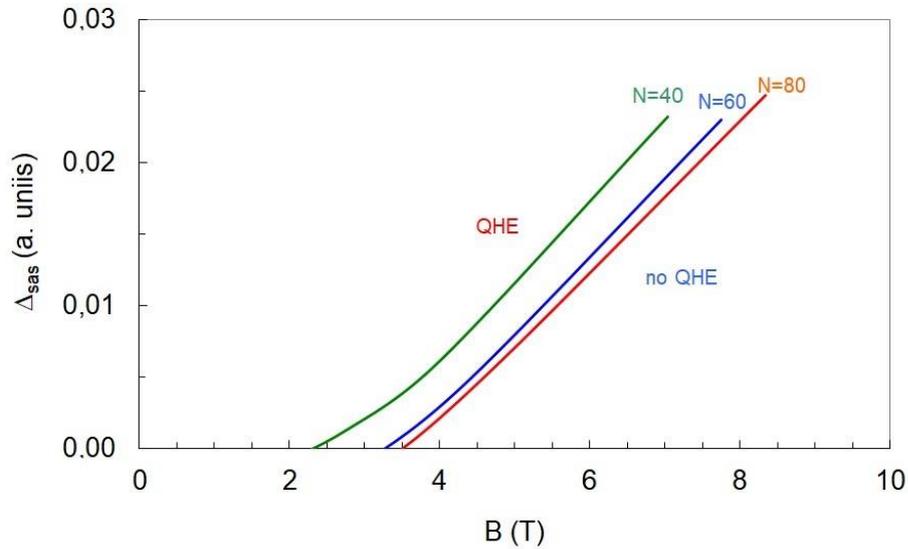

Figure 2. Dependence of the minimum tunnel splitting value on the magnetic field; $d$ =1.

Calculations were carried out at $\Delta_{sas} = 0$ for various values of $d$ in the range of values from 0.5 to 1. Results of these calculations are shown in Figure 3. It can be seen that as $d$ changes, the magnitude of the magnetic field at which QHS is stable, changes significantly, while the $d/L$ ratio varies slightly with a change in the magnetic field. From the results presented in Figures 2 and 3, it can be seen that QHS is always stable at $d/L < 0.7$ (the experimental results are $d/L <2$ [3] and $d/L <1.7$ [4]). This difference is apparently due to the influence of impurities and disorder on the electronic properties of the double layers. It is well known that impurities and disorders strongly affect the width of the Hall plateau in the regime of integer quantum Hall effect in single layers.

The influence of the impurity potential on the properties of the transition in QHS

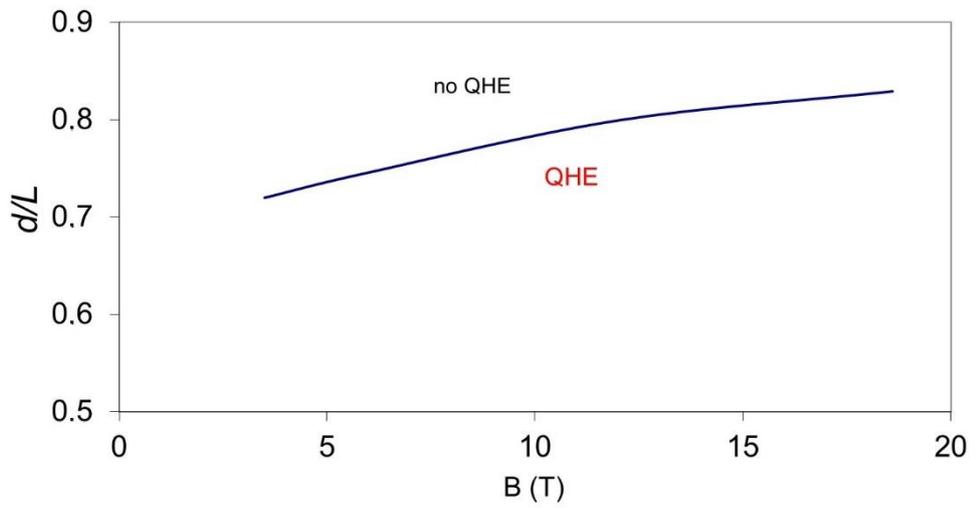

Figure 3. Dependence of the $d/L$ ratio on the magnetic field; $N = 80$, $\Delta_{sas} = 0$.

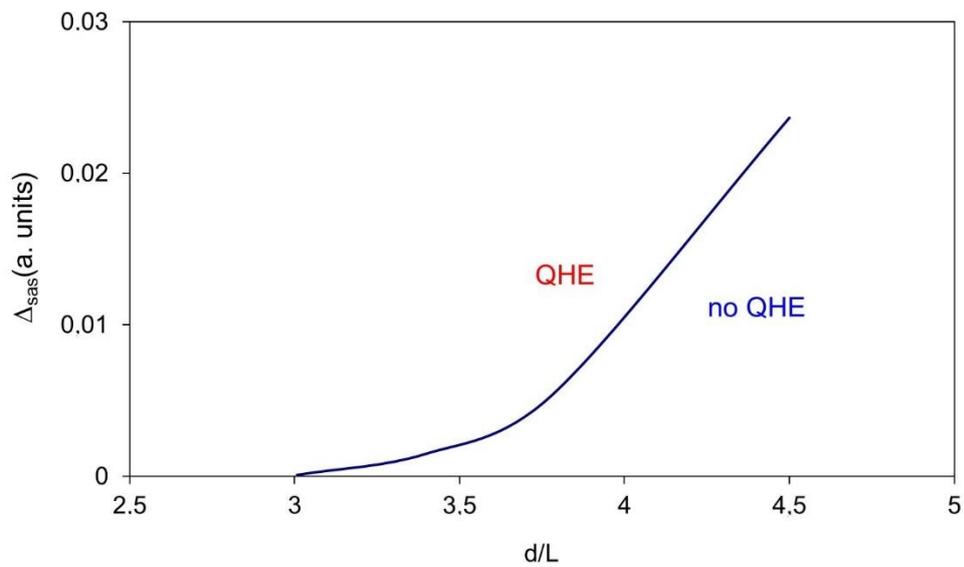

Figure 4. Phase diagram of the transition from QHS at $\nu =1$; $z_0 = 1$, $d =2.8$, $N=80$.

is studied. A point impurity with charge $z_0$ was taken as the impurity potential: $V(r) = -2z_0/\sqrt{r^2 + d^2/4}$. Density functional theory was used to calculate the critical value of the tunnel splitting, above which QHS is stable. From the results presented in Figure 4, it can be seen that QHS is always stable at $d/L < 3$. Note that as $z_0$ decreases, the critical value of $d/L$ decreases too.

## 4. Conclusion

The system of Kohn-Sham equations for spatially separated two-dimensional electrons in a strong magnetic field is self-consistently solved. Phase diagram of the transition in QHS in double quantum dots is plotted in coordinates of the tunnel splitting vs magnetic field strength at the Landau level filling factor $\nu = 1$. It is shown that QHS is stable at $d/L < 0.7$ in structures with zero tunnel splitting. It has been shown that the impurity potential has a significant effect on the stability condition of QHS at $\nu = 1$.

### Acknowledgements

This work was supported by the State Assignment of the Ministry of Science and Higher Education of the Russian Federation (project No. 0721-2020-0048).

E-mail: a_vas2002@mail.ru